\documentstyle[11pt,newpasp,twoside,epsf]{article}
\def\gr{$^\circ$}
\def\kms{km~s$^{-1}$}
\def\ml{M$_\odot$~yr$^{-1}$}
\def\nii{[N{\sc ii}]}

\def\edcomment#1{\iffalse\marginpar{\raggedright\sl#1\/}\else\relax\fi}
\reversemarginpar

\begin{document}
\title{Symbiotic Miras can do it}
\author{Romano L.M. Corradi}
\affil{Instituto de Astrof\'\i sica de Canarias, E-38200 La Laguna, Tenerife, 
Spain}
\author{Mario Livio}
\affil{Space Telescope Science Institute, 3700 San Martin Drive, Baltimore,
MD 21218, USA}
\author{Hugo E. Schwarz}
\affil{Nordic Optical Telescope, Apartado 474, E-38700 Sta. Cruz de La 
             Palma, Spain}
\author{Ulisse Munari}
\affil{Osservatorio Astronomico di Padova, Sede di Asiago, I-36032 Asiago, 
Italy}

\begin{abstract}
Symbiotic Miras give a nice practical demonstration of the formation of
bipolar and highly aspherical nebulae as a consequence of interaction in
detached binaries.  We believe that these binary systems are among the
most promising candidates for the progenitors of bipolar planetary nebulae.

We present a list of the optically extended nebulae known to date around
symbiotic Miras, and illustrate their properties using recent HST and ESO-NTT
\nii\ images of He~2-104 and He~2-147.
\end{abstract}

\section{Introduction}

Symbiotic Miras are interacting binary systems composed of a late AGB star
and a hot white dwarf. Even though the orbital periods of none of these
systems have been measured, they are thought to be of the order of 20-100 yr
(cf. Miko{\l}ajewska 1997).  Then interaction between the two stars does not
occur via Roche lobe overflow, but it is estimated that the white dwarf
accretes mass from the Mira wind at a rate of 10$^{-8}$~\ml\ or slightly
more. The accreted mass amounts to about 1$\%$ of the Mira wind; the rest
distributes around the binary system in a slowly expanding circumbinary
nebula.

Symbiotic Miras are a key case for the study of the effects on the AGB mass
loss caused by interaction in detached binaries. The Mira wind is partially
ionised by the white dwarf radiation, and thus it is more easily observable
than in single Miras. In addition, accretion of the Mira wind onto the white
dwarf causes nova--like eruptions and the consequent
production of fast winds which can last for decades (slow-novae).  These
winds interact and shape the bulk of the Mira wind which is not accreted. The
velocities and mass loss rate of the fast winds from the erupting hot
component of symbiotic stars are not very different from those produced by
the central stars of planetary nebulae (PNe). Thus, in symbiotic Miras not
only we can observe the effects of the binary interactions on the geometry of
the Mira wind, but also a ``simulation'' of its post--AGB evolution (PN)
under the action of a fast wind.

\section{The nebulae around symbiotic stars}

Our knowledge about the nebular environment of symbiotic stars has improved
significantly in recent years. Basic information about the morphology of the
innermost regions of these interacting binaries was obtained by means of
radio observations (see the list of references in Corradi et al. 1999a),
ground based long slit spectroscopic imaging (Solf 1983, 1984), and HST
imagery (Paresce \& Hack 1994; this paper).  In addition, ground--based CCD
imaging using specific narrow band filters allowed us to discover extended
ionised nebulae with sizes typical of planetary nebulae (up to more than one
parsec). In spite of the many similarities, it should be stressed that the
nebulae around symbiotic stars are not genuine PNe, since the gas is donated
by a star which is still on the AGB phase (pre-PN), while ionisation is
provided by the hot companion which has already lost its own PN.

To date, 8 optically extended nebulae are known around symbiotic Miras (for a
complete list of references of individual studies, see Corradi et
al. 1999a). Three of them have a bipolar shape (R~Aqr, BI~Cru, and He~2-104),
one is an inclined ring (He~2-147), three have a more irregular but markedly
aspherical shape (HM~Sge, V1016~Cyg and RX~Pup), and one is barely resolved
(H~1-36).

One of the most important things that we want to point out here, is that the
bipolar symbiotic nebulae have morphological and kinematical properties
(including high polar velocities of 200-300~\kms) which are strikingly
similar to those of some bipolar PNe (e.g. Hb~5 and NGC~6537, Corradi \&
Schwarz 1993).  This suggests a link between the two classes. The fact that
among symbiotic Miras half of the nebulae are bipolar/ring types, and all are
markedly aspherical, while among PNe only about 15$\%$ is bipolar, and some
20$\%$ is spherical, indicates that {\it interaction in detached binaries
favours the formation of bipolar and aspherical PNe}. This is confirmed by
theoretical modelling (Mastrodemos \& Morris 1999).

Moreover, several symbiotic Miras possess {\it multiple} nebulae which are
ascribed to the shaping action of fast winds set up during {\it recurrent}
outbursts.  This also offers a natural explanation for the origin of {\it
quadrupolar} PNe (Manchado, Stanghellini \& Guerrero 1996), whose formation
is difficult to understand in terms of interacting-winds evolution from single
stars.

Some of the properties of the extended nebulae around symbiotic Miras are
illustrated by the two cases below.

\subsection{He 2-104, the Southern Crab}

He~2-104 is a symbiotic star containing a Mira with a pulsational period of
400 days. Note that the presence of a Mira in this system is only
detected by the modulation of the IR luminosity, while the optical spectrum
does not show any sign of the cool stars, and appears as a typical spectrum
of a PN except for a core with peculiarly high (for PNe) densities.

We recently obtained an HST image of the multiple-bipolar nebula
around He~2-104 in the light of the \nii658.3~nm line. We present in
Fig.~1 the image as it appeared in an HST News Release on August 1999 (see
{\it http://oposite.stsci.edu/ pubinfo/pr/1999/32/}). This image nicely
resolves the inner bipolar nebula, which is a small scale reproduction
of the larger crab-like one. The inner nebula is strikingly similar to
the prototypical bipolar PN MyCn~18 (Sahai \& Trauger 1998),
suggesting a similar formation process.  At the distance of 800~pc
preferred by Schwarz et al. (1989), the inner and outer bipolar
nebulae of He~2-104 would have a kinematical age of 200 and 900~yr,
respectively.

\begin{figure}
\plotone{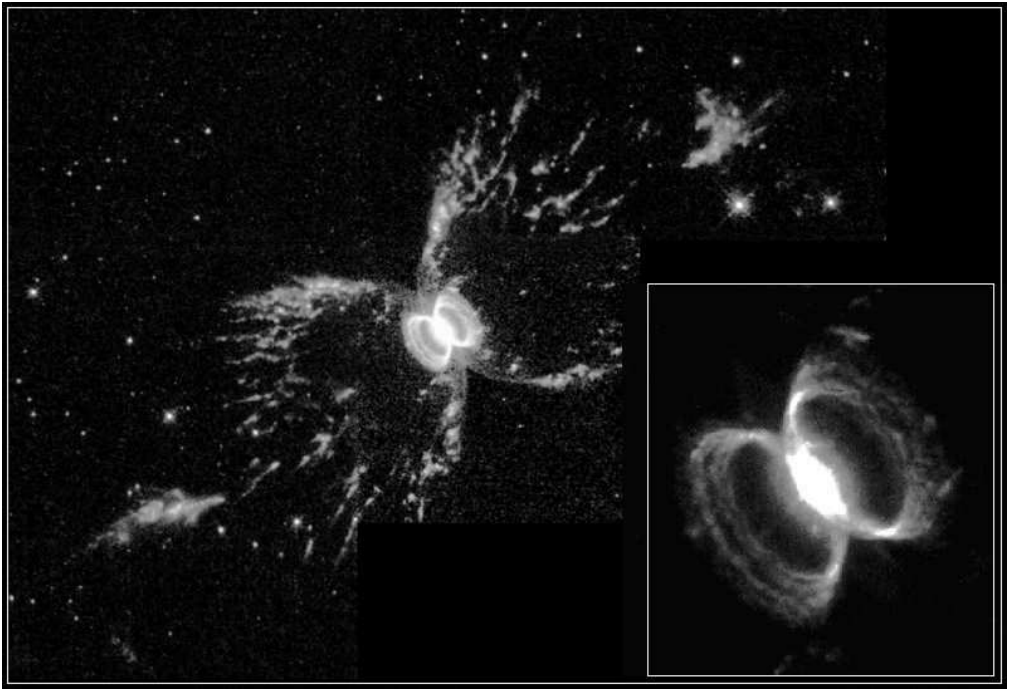}
\caption{HST \nii\ image of He~2-104. In the bottom-right box, a zoom of the
inner nebula.}
\end{figure}

\subsection{He 2-147}

According to Munari (1997), He~2--147 is a template case of a symbiotic nova
whose PN--like spectrum (which is typical of the outburst phase of these
systems) has ``retreated'' to the blue part of the spectrum leaving the red
region dominated by strong TiO bands and continuum from the Mira (which is
instead characteristic of the quiescent phase of symbiotic novae).  The
nebula around He~2--147 (Fig.~2) is a ring expanding with a velocity of
$\sim$100~\kms\ which is inclined at $\sim$ 55\gr\ to the line of sight
(Corradi et al. 1999b).  The kinematical age of the ring is of about 300~yr,
and its size is 0.05~pc.  Such a ring nebula can be considered an
``extreme bipolar'', in which material in the polar directions and at
intermediate latitudes (if any) is exceedingly tenuous, or has already
vanished into the surrounding space.
 
\begin{figure}
\plotone{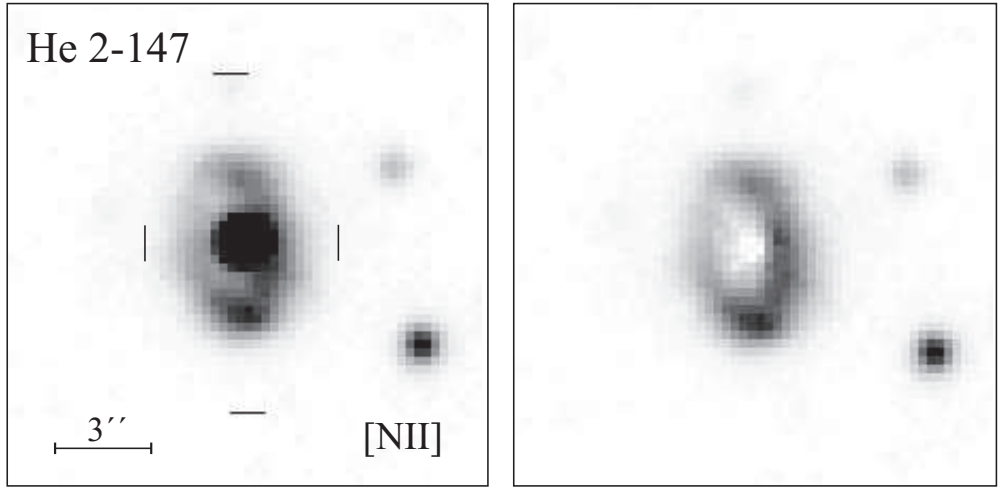}
\caption{NTT \nii\ image of He~2-147, before and after subtraction of the
central star emission.  North is at the top, East to the left (from Corradi
et al. 1999b).}
\end{figure}

\acknowledgements The work of RLMC is supported by a grant of the Spanish
DGES PB97-1435-C02-01.

\end{document}